\documentclass[epj]{svjour}

\def\be{\begin{equation}} 
\def\ee{\end{equation}} 
\def\bea{\begin{eqnarray}}
\def\ena{\end{eqnarray}}

\RequirePackage{graphicx}

\usepackage{cite} 
\usepackage{amsmath}
\usepackage{hyperref}
\usepackage{amsmath,amssymb}
\usepackage{xcolor}

\usepackage{graphics}

\begin{document}
\title{Anisotropic strange quark stars with a non-linear equation-of-state}

\author{
Il{\'i}dio Lopes \inst{1} 
\thanks{E-mail: \href{mailto:ilidio.lopes@tecnico.ulisboa.pt}{\nolinkurl{ilidio.lopes@tecnico.ulisboa.pt}} }
\and
Grigoris Panotopoulos \inst{1} 
\thanks{E-mail: \href{mailto:grigorios.panotopoulos@tecnico.ulisboa.pt}{\nolinkurl{grigorios.panotopoulos@tecnico.ulisboa.pt}} }
\and
\'Angel Rinc\'on \inst{2}
\thanks{E-mail: \href{mailto:angel.rincon@pucv.cl}{\nolinkurl{angel.rincon@pucv.cl}} }
}                     
%
%
\institute{ 
Centro de Astrof\'{\i}sica e Gravita{\c c}{\~a}o, Departamento de F{\'i}sica, Instituto Superior T\'ecnico-IST,
\\
Universidade de Lisboa-UL, Av. Rovisco Pais, 1049-001 Lisboa, Portugal.
\and
Instituto de F\'isica, Pontificia Universidad Cat\'olica de Valpara\'iso, Avenida Brasil 2950, Casilla 4059, Valpara\'iso, Chile.
}

\date{Received: date / Revised version: date}
%
\abstract{
We obtain an exact analytical solution to Einstein's field equations assuming a non-linear equation-of-state and a particular mass function. Our solution describes the interior of anisotropic color flavor locked strange quark stars. All energy conditions are fulfilled, and therefore the solution obtained here is a realistic solution within General Relativity.
The mass-to-radius profile is obtained, and the compactness of the star is computed.
}

\maketitle

\section{Introduction}

Unlikely many other forms of matter, compact objects \cite{textbook,review1,review2}, which are formed at the end stages of stellar evolution, are unique probes to study the properties of matter under exceptionally extreme conditions. The matter inside such objects is characterized by ultra-high matter densities for which the usual classical description of stellar plasmas in terms of non-relativistic Newtonian fluids is inadequate. Therefore, such very dense compact objects are relativistic and as such, they are only properly described within the framework of Einstein's General Relativity (GR) \cite{GR}.  

\smallskip

Soon after the discovery of the neutron by James Chadwick \cite{neutron1,neutron2}, Baade and Zwicky in a seminal work published in 1934 predicted the existence of neutron stars (NSs) as the endpoint of supernovae \cite{baade}. The indisputable observational proof came one year after the discovery of pulsars in 1967 \cite{Hewish68}, following  the insightful prediction of Pacini \cite{Pacini1967} that pulsars are rapid rotating neutron stars, which can be found in the central region of supernova remnants, as the ones discovered at the core of the Crab and Vela supernovas \cite{chamel}. 

\smallskip

Moreover, neutron stars are exciting stellar astrophysical objects since the understanding of their properties and their observed complex phenomena requires bringing together several different scientific disciplines and lines of research, such as nuclear particle physics, astrophysics and gravitational physics. As they are the denser objects in the Universe after black holes, thanks to their extreme conditions which cannot be reproduced on earth-based experiments, they constitute an excellent cosmic laboratory to study and constrain non-conventional physics and alternative theories of gravity.

\smallskip

Strange quark stars \cite{SS1,SS2,SS3,SS4,SS5,SS6}, at the moment hypothetical objects, can be viewed as ultra-compact NSs. Since quark matter is by assumption absolutely stable, it may be the true ground state of hadronic matter \cite{witten,farhi}, and therefore this new class of relativistic compact objects has been proposed as an alternative to typical NSs. In fact strange quark stars may explain the observed super-luminous supernovae \cite{SL1,SL2}, which occur in about one out of every 1000 supernovae explosions, and which are more than 100 times more luminous than regular supernovae. One plausible explanation is that since quark stars are much more stable than NSs, they could explain the origin of the huge amount of energy released in super-luminous supernovae. Many works have been recently proposed to validate its existence in different astrophysical scenarios \cite{PL18a,Mukhopadhyay16}. 

\smallskip

In studies of compact relativistic astrophysical objects the authors usually focus on stars made of an isotropic fluid, where the radial pressure $p_r$ equals the tangential pressure $p_T$. However, celestial bodies are not always made of isotropic fluid only. In fact under certain conditions the fluid can become anisotropic. The review article of Ruderman~\cite{paper1} mentioned for the first time  such a possibility:  this author makes the observation that relativistic particle interactions in a very dense nuclear matter medium could lead to the formation of anisotropies. The study on  anisotropies in relativistic stars has received a boost by the subsequent work of \cite{paper2}. Indeed, anisotropies can arise in many scenarios  of a dense matter medium, like phase transitions \cite{paper3}, pion condensation \cite{paper4}, or in presence of type $3A$ super-fluid \cite{paper5}. See also \cite{Ref_Extra_1,Ref_Extra_2,Ref_Extra_3} for more recent works on the topic, and references therein. In these works relativistic models of anisotropic quark stars were studied, and the energy conditions were fulfilled. In particular, in \cite{Ref_Extra_1} an exact analytical solution was obtained, in \cite{Ref_Extra_2} an attempt was made to find  a singularity free solution to Einstein's field equations, and in \cite{Ref_Extra_3} the Homotopy Perturbation Method was employed, which is a tool that facilitates to tackle Einstein's field equations. What is more, alternative approaches have been considered to incorporate anisotropies into known isotropic solutions \cite{Gabbanelli:2018bhs,Ovalle:2017fgl,Ovalle:2017wqi}.

\smallskip

In \cite{anisotropy} a new exact solution was obtained assuming a particular mass function. The solution obtained in that work described the interior of an anisotropic strange quark star with a linear equation-of-state (EoS), $p_r = n (p-\rho_s)$, with $\rho_s$ being the surface energy density, and $n$ a number that takes values in the range $0 \leq n \leq 1$. Another exact analytical solution for relativistic anisotropic quark stars was obtained in \cite{Ref_Extra_1}, assuming a linear EoS and a particular density profile instead of a particular mass function.
It is known, however, that other more refine and sophisticated quark EoSs exist in the literature, such as color superconductivity \cite{CFL1,CFL2,CFL3}, models that incorporate a chiral symmetry breaking \cite{NJL1,NJL2} or models based on perturbative QCD studies \cite{refine1,art}, and others \cite{refine2,refine3}. All these EoSs lead to stars the properties of which vary in stiffness and compactness.

\smallskip

In the mass-radius plane the highest star mass that a given EoS can support crucially depends on the equation-of-state, and soft equations of state predict lower highest masses \cite{paper}. The observed massive pulsars PSR J1614-2230 with a mass $(1.97 \pm 0.04) M_{\odot}$ \cite{shapiro} and PSR J0348-0432 with a mass $(2.01 \pm 0.04) M_{\odot}$ \cite{antoniadis} have put a stringent constrain on the EoS. Several EoSs predict maximum star masses well below $2 M_{\odot}$, and therefore they must be ruled out. The linear EoS in particular, considered in \cite{anisotropy}, can support a highest star mass well below the 2 solar mass bound.

\smallskip

Therefore, in the present article we propose to investigate the properties of anisotropic strange quark stars adopting a slightly more complicated, non-linear EoS obtained in the framework of QCD superconductivity \cite{CFL1,CFL2,CFL3}. Although the mass function will be assumed to be the same as the one considered in \cite{anisotropy}, the new EoS considered here leads to stars with different properties.

\smallskip

The plan of our work is the following: In the next section we present the structure equations describing hydrostatic equilibrium. In section 3 we obtain the exact analytical solution, and we discuss its significance for the properties of the stars. Finally we conclude in the last section. We adopt the mostly positive metric signature, $(-,+,+,+)$, as well as natural units in which both the speed of light in vacuum and the reduced Planck constant are set to unity, $c=1=\hbar$. Thus, all dimensionful quantities are measured in GeV, and we make use of the following conversion rules $1 \ \text{m} = 5.068 \times 10^{15} \ \text{GeV}^{-1}$ and $1 \ \text{kg} = 5.610 \times 10^{26} \ \text{GeV}$ \cite{guth}.

\section{Structure equations}

The model is described by the action
\begin{equation}
S \equiv S_G + S_M,
\end{equation}
where the gravity part $S_G$ is given by the usual Einstein-Hilbert term, while the matter contribution corresponds to a perfect fluid with energy density $\rho$, radial pressure $p_r$, tangential pressure $p_T$, and a certain equation-of-state $F(p_r,\rho)=0$. Varying with respect to the metric tensor we obtain Einstein's field equations without a cosmological constant
\begin{equation}
G_{\mu \nu} = R_{\mu \nu}-\frac{1}{2} R g_{\mu \nu}  = 8 \pi T_{\mu \nu}
\end{equation}
where we have set Newton's constant $G$ equal to unity, and the stress-energy tensor is given by
\begin{equation}
T_\mu ^\nu = \text{diag}(-\rho, p_r, p_T, p_T)
\end{equation}
where $p_r \neq p_T$, and we define the anisotropy to be $\Pi \equiv p_T-p_r$.

For the exterior problem, $r > R$, with $R$ being the radius of the star, 
where $T_{\mu \nu}$ vanishes, we seek static spherically symmetric solutions of the form
\begin{equation}
\mathrm{d}s^2 = -f(r) \mathrm{d}t^2 + g(r)  \mathrm{d}r^2 + r^2 (\mathrm{d} \theta^2 + \sin^2{\theta} \mathrm{d} \phi^2)
\end{equation}
The solution to the exterior problem is of course the well-known Schwarzschild solution \cite{SBH}
\begin{equation}
f(r) = g(r)^{-1} = 1 - \frac{2M}{r}
\end{equation}
with $M$ being the mass of the star.

To obtain the interior solutions, $r < R$, we need to solve the field equations in the presence of the anisotropic fluid. As usual for non-rotating objects we make the common ansatz
\begin{equation}
\mathrm{d}s^2 = -e^{\nu(r)} \mathrm{d}t^2 + A(r) \mathrm{d}r^2 + r^2 (\mathrm{d} \theta^2 + \sin^2{\theta} \mathrm{d} \phi^2)
\end{equation}
and we set for convenience
\begin{equation}
A(r)^{-1} = 1 - \frac{2m(r)}{r}
\end{equation}
similarly to the exterior solution, where $m(r)$ is the mass function.
Therefore, in total there are four unknown functions, $m(r), \nu(r), p_r(r), \Pi(r)$, satisfying the following system of coupled differential equations \cite{anisotropy}
\begin{eqnarray}
m' & = & 4 \pi \rho r^2 
\label{mprime}
\\
p_r' & = & -(p_r+\rho) \frac{m + 4 \pi r^3 p_r}{r (r-2 m)} + \frac{2}{r} \Pi
\label{prprime}
 \\
\nu' & = & 2 \frac{m + 4 \pi r^3 p_r}{r (r-2 m)}
\label{nuprime}
\end{eqnarray}
where $m$, $\rho$, $p_r$, $\Pi$ and $\nu$ are functions of 
the radial coordinate $r$, and the prime denotes differentiation with respect $r$.
For isotropic stars, $\Pi(r)=0$, we recover the usual Tolman-Oppenheimer-Volkoff equations \cite{OV,tolman}. Finally, the differential equations are supplemented with the appropriate initial conditions at the center of the star, which are the following
\begin{eqnarray}
p_r(0) & = & p_c \\
m(0) & = & 0 
\end{eqnarray}
with $p_c$ being the central pressure. Upon matching the two solutions on the surface of the star, the following conditions must be satisfied
\begin{eqnarray}
p_r(R) & = & 0 \\
p_T(R) & = & 0 \\
m(R) & = & M \\
e^{\nu(R)} & = & 1 - \frac{2M}{R}
\end{eqnarray}
The first and the second conditions allow us to determine the radius of the star, while the third one allows us to compute its mass. 

\begin{figure}[ht!]
\centering
\includegraphics[width=0.8\linewidth]{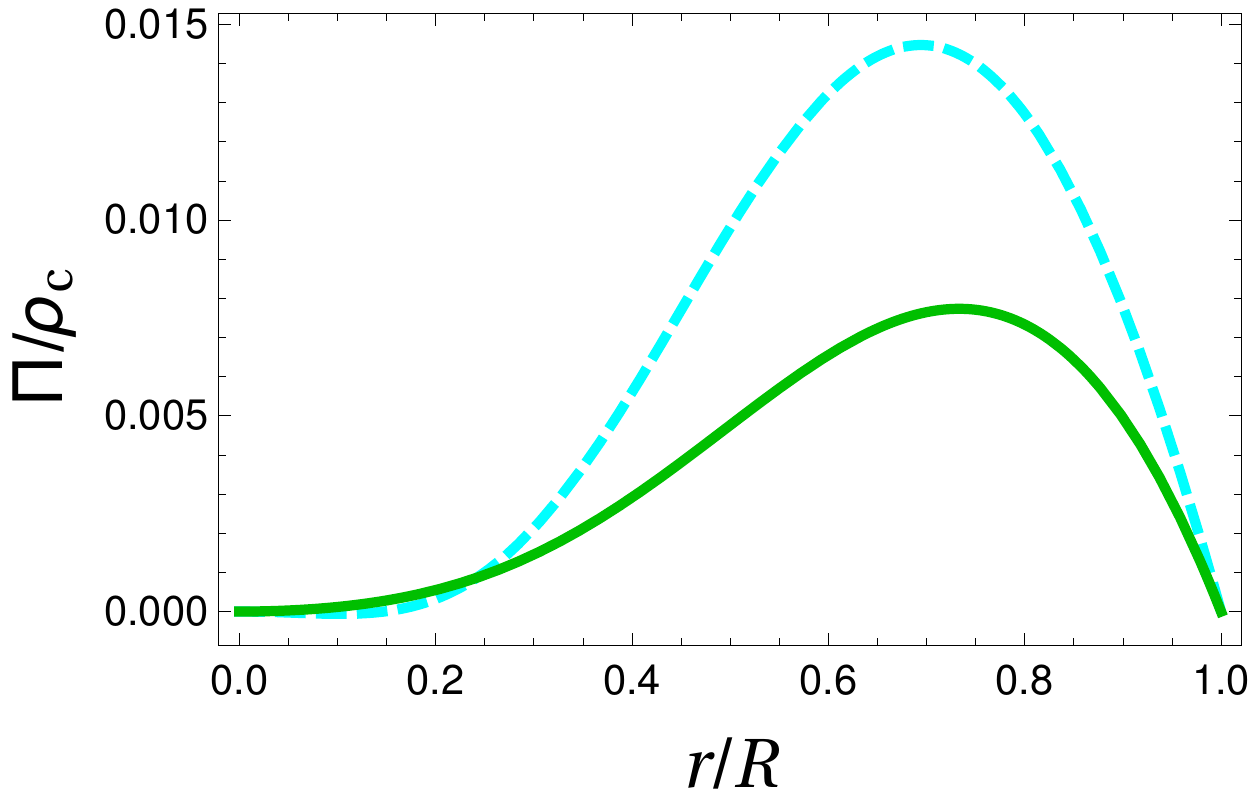}
\caption{
Anisotropy versus radial distance (both normalized) for both models, CFL 19 in green and CFL 10 in cyan, for $\tilde{b}=20$. The corresponding value of $\tilde{a}$ as well as the properties of the stars are shown in the tables~\ref{table:firstset} and \ref{table:secondset}. 
}
\label{fig:1} 
\end{figure}

\begin{figure*}[ht]
\centering
\includegraphics[width=0.49\textwidth]{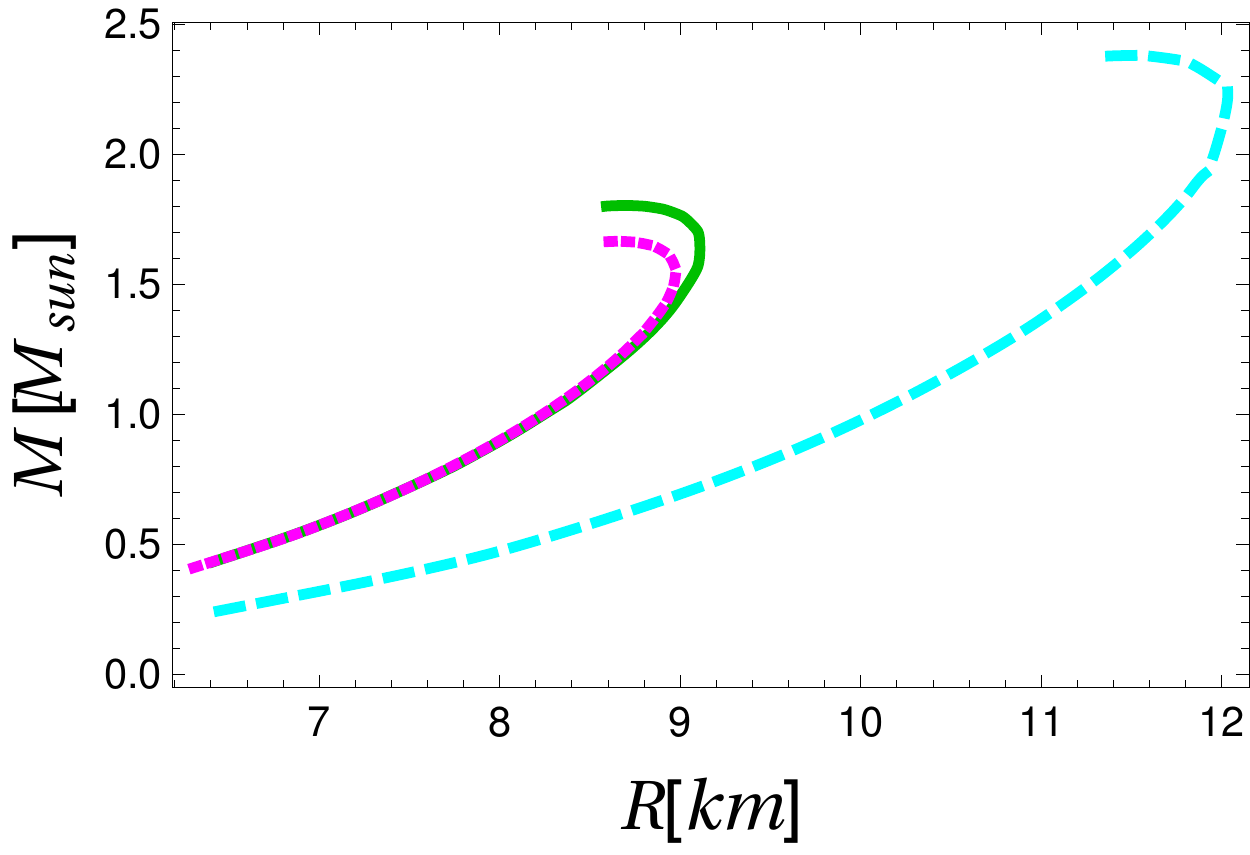}   \
\includegraphics[width=0.49\textwidth]{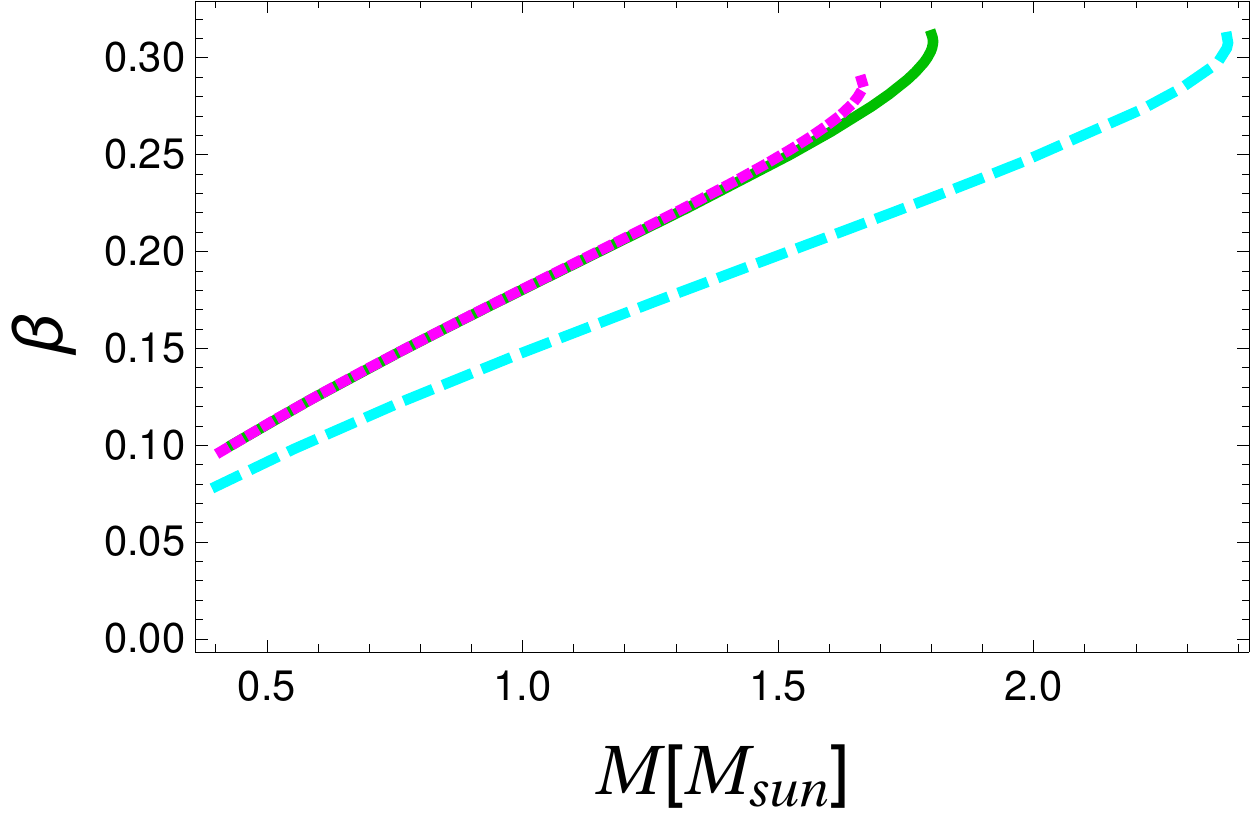}   \
\caption{
		{\bf Left panel:} 
		Mass-to-radius profiles: mass in solar masses vs radius in km.
		{\bf Right panel:} 
		Compactness $\beta = M/R$ vs mass of the star (in solar masses). Colours are as follows: Anisotropic CFL 19 (green), anisotropic CFL 10 (cyan) and isotropic CFL 19 (magenta).
}
\label{fig:2}
\end{figure*}

\section{Exact solution}

To solve the structure equations we need to specify the source first. In the present work, we adopt the EoS obtained in the framework of color superconductivity due to its mathematical simplicity, as it is always desirable to work with an analytic function. It certainly would be interesting to see precisely how other EoSs would affect the results of the present work, and we hope to be able to address that issue in a future work.

We model matter inside the star as a relativistic gas of de-confined quarks, and we consider a non-linear extension of the simplest version of the MIT bag model \cite{bagmodel1,bagmodel2}
\begin{equation}
p = \frac{1}{3} (\rho - 4B)
\end{equation}
with $B$ being the bag constant. Within the framework of color superconductivity one obtains a slightly more complicated EoS, which is given by \cite{CFL1,CFL2,CFL3}
\begin{equation}
p_r = \frac{1}{3} (\rho - 4B) - \frac{9 \alpha}{\pi^2} \mu^2 \label{pressure}
\end{equation}
\begin{equation}
\mu^2 = - \alpha + \sqrt{\alpha^2 + (4/9) \pi^2 (\rho - B)}
\end{equation}
where the parameter $\alpha$ is given by
\begin{equation}
\alpha = \frac{2 \Delta^2}{3} - \frac{m_s^2}{6}
\end{equation}
with $m_s$ being the mass of the strange quark, and $\Delta$ the energy gap. Clearly, when $\alpha \rightarrow 0$ one recovers the linear "radiation plus constant" EoS.

In the framework of color superconductivity the equation-of-state of quark matter is characterized by three free parameters, $B, m_s, \Delta$. In total there are 19 viable models, out of which 8 admit a highest star mass larger than 2 solar masses \cite{CFL3}. In the discussion to follow, we choose for our study here two fiducial models (number 19 and number 10) of \cite{CFL3}, where 
\begin{eqnarray}
\Delta & = & 150  \ \text{MeV} \\
m_s & = & 0 \\
B & = & 140 \ \text{MeV}  \ \text{fm}^{-3}
\end{eqnarray}
and $M_{max}=1.67~M_{\odot}$ for the isotropic  model CFL 19, and
\begin{eqnarray}
\Delta & = & 150  \ \text{MeV} \\
m_s & = & 150 \ \text{MeV} \\
B & = & 80 \ \text{MeV}  \ \text{fm}^{-3}
\end{eqnarray}
and $M_{max}=2.20~M_{\odot}$ for the isotropic model CFL 10. Note that the numerical value of the bag constant considered here is different than the values
considered in \cite{Ref_Extra_1,Ref_Extra_2,Ref_Extra_3}, although one of their values ($83~MeV fm^{-3}$) is very close to the value of CFL 10.

\smallskip

To find a tractable exact solution one may assume a particular mass function, as was done in \cite{anisotropy}, or a particular density profile, as was done in \cite{Ref_Extra_1}. Clearly both approaches are acceptable, and they may be equally used. We find it slightly more convenient, though, to assume a mass function and obtain the energy density upon differentiation with respect to $r$. In previous works \cite{anisotropy,older1,older2} the authors worked with the following mass function
\begin{equation}
m(r) = \frac{b r^3}{2(1+a r^2)}
\end{equation}
which will be adopted here as well, characterized by two parameters, $a,b$, with dimensions of $\text{length}^{-2}$.
This choice for the mass function, using equation (\ref{mprime}), 
leads to the following expression for the energy density
\begin{equation}
\rho(r) = \frac{b (3+a r^2)}{(1+ar^2)^2}
\end{equation}
with central value $\rho_c=\rho(0)=3b$, which has the appealing feature of being a monotonically decreasing function of the radial coordinate $r$.
The EoS allows us to compute the radial pressure using equation (\ref{pressure}), and then the anisotropy can be computed using the second structure equation (\ref{prprime}). Finally, $\nu(r)$ can be obtained by the integration of equation (\ref{nuprime}) in a straightforward manner since all functions are now known, although the resulting expression is quite complicated. Luckily, in order to compute the radius and the mass of the star, the function metric $\nu(r)$ is not necessary.

We wish to stress at this point that as far as the technique is concerned, the exact analytical solution obtained here is not better than the ones found in \cite{Ref_Extra_1,anisotropy}. The improvement that we propose here lies on the choice of the EoS, which corresponds to a physically different description of the interior solution of the stars.

\subsection{Numerical results}

Here we analyze the consequences of the solution. We introduce dimensionless quantities $\tilde{a},\tilde{b}$ as follows
\begin{eqnarray}
a & = & \frac{\tilde{a}}{r_0^2} \\
b & = & \frac{\tilde{b}}{r_0^2} 
\end{eqnarray}
where $r_0=43.245~km$. For a given $b$ the two conditions $p_r(r;a,b)=0=p_T(r;a,b)$ allow us to determine the radius of the star as well as the parameter $a$. Therefore, only $b$ is a free parameter, while $a$ is not an independent parameter.

In Tables~\ref{table:firstset}, \ref{table:secondset} and \ref{table:thirdset} we show the star radii (in km), the masses (in solar masses) and the compactness for several values of $b$ both for the non-linear and the linear EoSs for comparison. For the same bag constant the non-linear EoS predicts larger and more massive stars than the linear EoS.

In Fig.~\ref{fig:1} we show the anisotropy $\Pi$ as a function of the radial coordinate $r$ for both models CFL 19 and CFL 10. We use dimensionless quantities, $\Pi/\rho_c$ and $r/R$, respectively. We see that the behaviour is qualitatively very similar to the behaviour of the anisotropy of the solution found in \cite{Ref_Extra_1}, see Fig.~1 of that work.

In Fig.~\ref{fig:2} we show the mass-to-radius profile (left panel) as well as the compactness of the star $\beta=M/R$ versus mass of the star (right panel) for both models CFL 19 and CFL 10. The profile corresponding to the isotropic star CFL 19 is also shown for comparison. It is worth noticing that the maximum mass of the anisotropic star is considerably higher compared to its isotropic counterpart, $M_{max} \simeq 1.8~M_{\odot}$ versus $1.67~M_{\odot}$, \cite{CFL3}, although it still does not intersect the massive pulsars \cite{shapiro,antoniadis} band. The maximum radius, too, of the anisotropic star is larger than the maximum radius of its isotropic counterpart, a fact already observed in \cite{Ref_Extra_1}.

\begin{table*}
\centering
  \caption{Star properties for $\alpha \neq 0$ (model CFL 19)}
  \begin{tabular}{ccccc}
  \hline
$\tilde{b}$ & $\tilde{a}$ & R [km] & M [$M_{\odot}$] & $\beta$ \\
\hline
13 &  7.54 & 8.45 & 1.09 & 0.19  \\
18 & 13.24 & 9.06 & 1.52 & 0.25  \\
20 & 15.62 & 9.11 & 1.60 & 0.26  \\
25 & 21.76 & 9.08 & 1.72 & 0.28  \\
30 & 28.12 & 8.99 & 1.77 & 0.29 \\
\hline
\end{tabular}
\label{table:firstset}
\end{table*}

\begin{table*}
\centering
  \caption{Star properties for $\alpha \neq 0$ (model CFL 10)}
  \begin{tabular}{ccccc}
  \hline
$\tilde{b}$ & $\tilde{a}$ & R [km] & M [$M_{\odot}$] & $\beta$ \\
\hline
13 & 10.84 & 12.03 & 2.21 & 0.27  \\
18 & 17.18 & 11.84 & 2.35 & 0.29  \\
20 & 19.79 & 11.74 & 2.37 & 0.30  \\
25 & 26.48 & 11.50 & 2.38 & 0.31  \\
30 & 33.33 & 11.30 & 2.37 & 0.31  \\ 
\hline
\end{tabular}
\label{table:secondset}
\end{table*}

\begin{table*}
\centering
  \caption{Star properties for $\alpha = 0$ (Bag constant same as in previous table)}
  \begin{tabular}{ccccc}
  \hline
$\tilde{b}$ & $\tilde{a}$ & R [km] & M [$M_{\odot}$] & $\beta$ \\
\hline
13 & 10.62 & 9.51 & 1.33 & 0.21  \\
18 & 17.27 & 9.70 & 1.58 & 0.24  \\
20 & 20.02 & 9.68 & 1.63 & 0.25  \\
25 & 27.10 & 9.58 & 1.70 & 0.26  \\
30 & 34.38 & 9.45 & 1.72 & 0.27  \\ 
\hline
\end{tabular}
\label{table:thirdset}
\end{table*}

\subsection{Energy conditions}

The obtained solution should be able to describe realistic astrophysical configurations.
Therefore, in this subsection as a final check we investigate if the energy conditions 
are fulfilled or not. We require that \cite{Ref_Extra_1,Ref_Extra_2,Ref_Extra_3}
\begin{equation}
\rho \geq 0
\end{equation}
\begin{equation}
\rho + p_{r,T}  \geq 0
\end{equation}
\begin{equation}
\rho - p_{r,T}  \geq  0
\end{equation}
\begin{equation}
E_- \equiv \rho - p_r -2 p_T \geq 0
\end{equation}
\begin{equation}
E_+ \equiv \rho + p_r +2 p_T \geq 0
\end{equation}
Given the solution obtained in the previous section, it is obvious that the first condition is automatically satisfied. First we shall check if the fourth and the fifth conditions are satisfied for both models considered in this work, namely model CFL 19 and CFL 10. In the last figure we show both $E_-, E_+$ versus normalized radius coordinate, $r/R$, for both models and for $\tilde{b}=20$. Clearly, they are positive throughout the star. We have checked that $E_-, E_+$ are positive for the rest of the values of $\tilde{b}$ as well, although we have not showed them all in the figure. What is more, it is easy to check that $p_r,p_T$ are positive throughout the star (which is confirmed by the fact that $E_+ > E_-$), and therefore the rest of the energy conditions are satisfied as well. We thus conclude that the exact analytical solutions obtained in the present work are realistic GR solutions, which are able to describe realistic astrophysical configurations.

\begin{figure}[ht!]
\centering
\includegraphics[width=0.8\linewidth]{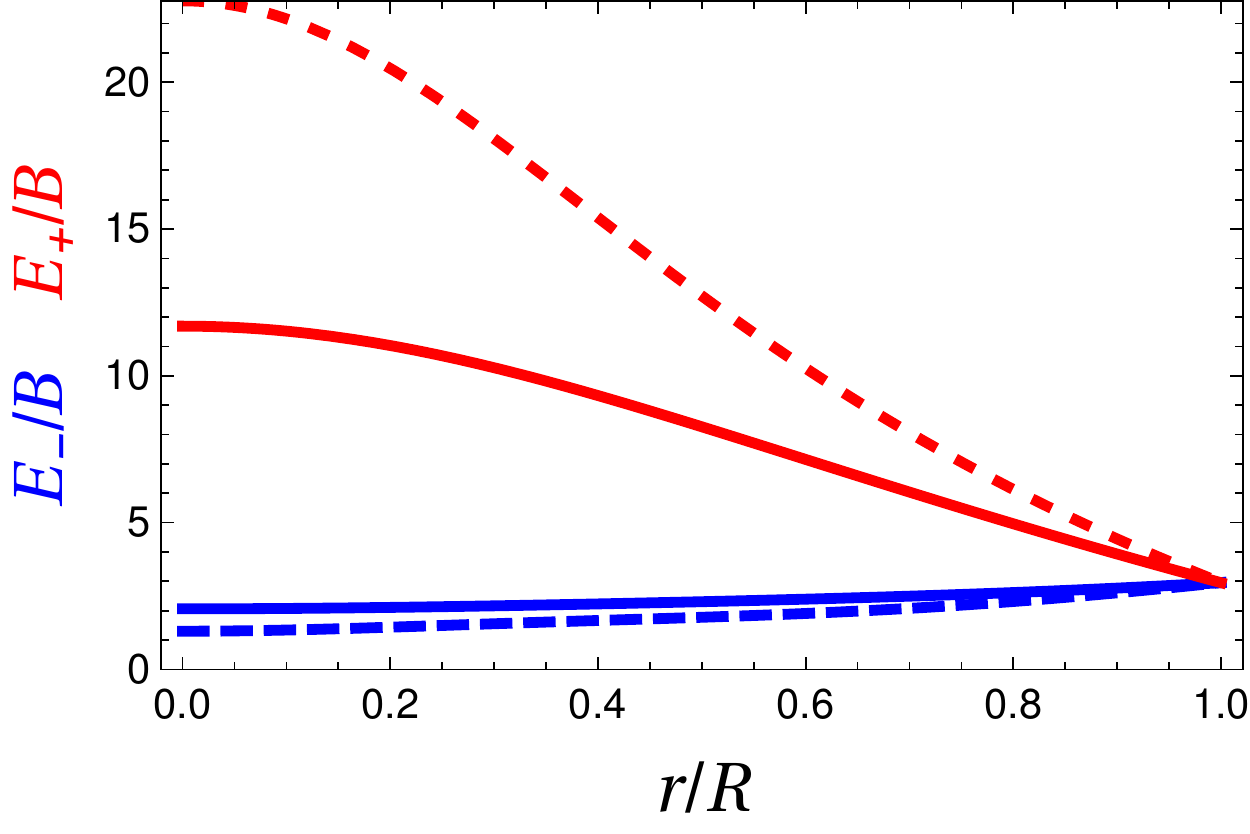}
\caption{
Energy conditions $E_-,E_+$ (see text) for models CFL 19 (solid curves) and CFL 10 (dashed curves) for $\tilde{b}=20$. Blue lines correspond to $E_-$ and red lines correspond to $E_+$.
}
\label{fig:3} 
\end{figure}

\section{Conclusions}

In summary, in the present work we have obtained exact analytical interior solutions for non-rotating anisotropic relativistic stars. We have assumed a non-linear equation-of-state, which describes color flavor locked strange quark stars. Given the complexity of the structure equations, the starting point was to assume a particular mass function characterized by two parameters $a,b$. After that, the energy density, the radial pressure and the anisotropy were computed in a straightforward manner using the structure equations. We have computed the star radii and masses for several values of $a,b$ both for the non-linear EoS and for the linear one for comparison reasons. The mass-to-radius profile has been obtained, and the compactness of the star has been computed. Finally, all energy conditions are satisfied.


\section*{Acknowlegements}

We wish to thank the anonymous reviewer for useful suggestions.
The authors I.~L. and G.~P. thank the Funda\c c\~ao para a Ci\^encia e Tecnologia (FCT), Portugal, for the financial support to the Center for Astrophysics and Gravitation-CENTRA,  Instituto Superior T\'ecnico, Universidade de Lisboa, through the Grant No. UID/FIS/00099/2013. 
The author \'A.~R. acknowledges DI-VRIEA for financial support through Proyecto Postdoctorado 2019 VRIEA-PUCV.


\end{document}